\documentstyle[preprint,aps]{revtex}

\begin{document}

\def\la{\mathrel{\mathpalette\fun <}}
\def\fun#1#2{\lower3.6pt\vbox{\baselineskip0pt\lineskip.9pt
  \ialign{$\mathsurround=0pt#1\hfil##\hfil$\crcr#2\crcr\sim\crcr}}}

\title{WKB Wave Functions with the Induced Gravity~Theory}

\author{Zong-Hong Zhu\thanks{e-mail address: zhuzh@class1.bao.ac.cn} 
	and Li Cao\thanks{e-mail address: caoli@nova.bao.ac.cn}}
\address{Beijing Astronomical Observatory,
		Chinese Academy of Sciences, Beijing 100012, China\\
	National Astronomical Observatories,
		Chinese Academy of Sciences, Beijing 100012, China
	}

\maketitle

\begin{abstract}

The Wheeler-DeWitt equation for the induced gravity theory is constructed in
the minisuperspace approximation, and then solved using the WKB method
under three types of boundary condition proposed respectively by 
Hartle \& Hawking (``no boundary''), Linde and Vilenkin (``tunneling 
from nothing''). 
It is found that no matter how the gravitational and cosmological 
``constants'' vary in the classical models, they will 
acquire constant values when the universe comes from quantum creation, 
and that, in particular, the resulting tunneling wave function under the
Linde or Vilenkin  boundary condition reaches its maximum value  if the 
cosmological constant vanishes.\\


\end{abstract}

\newpage

\section{Introduction}

It is known that the Hartle-Hawking quantum cosmology is based on 
Einstein's general relativity.
Although the latter is extremely successful at describing the observable 
universe, it doesn't fully incorporate Mach's principle, which demands
that spacetime is determined entirely by background matter fields and
physical laws.
Hence the other type of theories, notably the Brans-Dicke theory\cite{bra61}
and the induced gravity theory\cite{zee79}, in which the gravitational and 
cosmological ``constants'' can result from a scalar field.
In such a gravity theory, both the the gravitational and cosmological 
``constants'' are dynamical and time-dependent quantities.
Observationally, there exist a number of experimental constraints on the
time variation of the gravitational ``constant'', $G$\cite{kra92gil97}, 
of which the tightest bound, 
$|\dot{G}/G| = (0.6\pm 2.0)\times 10^{-12} {\rm yr}^{-1} $,
was found by Thorsett\cite{tho96} using Bayesian statistical techniques on
the measurements of the masses of young and old neutron stars in pulsar
binaries.
The upper limits on the time variation of the cosmological ``constant'', 
$\Lambda$, can be deduced from number counts of faint galaxies\cite{yos92},
statistical properties of gravitational lensing\cite{blo96},
structure formation\cite{via97} and other ways. All results indicate
an almost constant $\Lambda$. Moreover, many attempts have been made
in order to develop a plausible model in which the cosmological constant 
$\Lambda$ is set to be precisely zero\cite{wei89,car92}.
So a critical problem is how the universe acquires almost constant values 
for $G$ and $\Lambda$, and especially, a vanishing value for the latter.
Because quantum cosmology (for an elegant review, see Ref.~\cite{hal91}) 
could, with no more than physical laws, provide a scheme which explains
the present universe is what it is, we should endeavor to solve the problems
mentioned above within the framework of such theory.
In this context, We have considered the Brans-Dicke theory in previous
papers\cite{zhu909298}, here we consider the induced gravity theory.

The first consideration of the quantum cosmology based on the induced
gravity theory was given by Mo and Fang using the path integral 
technique\cite{mo88}.
In this {\it letter}, we try to tackle the problem using the canonical
quantization method, concentrating particularly on the time variation
of $G$ and $\Lambda$.
The Wheeler-Dewitt equation (WDWE) is constructed  in the minisuperspace 
approximation and the wave functions of the universe are obtained using
three kinds of boundary condition, that are proposed by Hartle \&
Hawking\cite{har83}, Linde\cite{lin84} and Vilenkin\cite{vil86} respectively.
We shall show that  no matter how $G$ and $\Lambda$ vary in the classical
models, they will acquire constant values when the universe comes from 
quantum creation.
Moreover, the amplitude of the resulting wave function under the Vilenkin or
Linde boundary condition sharply peaks around the classical trajectory only
for a vanishing cosmological constant.

\section{The Wheeler-Dewitt equation for the Induced\\ Gravity Theory}

The action of induced gravity is\cite{zee79,mo88}
\begin{equation}
\label{action}
S = \int d^4x \sqrt{-g} \left[\frac{1}{2}\epsilon \varphi^2R -  
	\frac{1}{2}g^{\mu\nu}\varphi_{,\mu}\varphi_{,\nu} - V(\varphi)\right] + ... \,\,\,\,.
\end{equation}
So the scalar field $\varphi$ induces a universe where the gravitational 
and cosmological ``constants'' are given simutaneously by
\begin{equation}
\label{constants}
\begin{array}{rl}
(16\pi G_{\rm ind})^{-1} & = \frac{1}{2}\epsilon\varphi^2  \,\,,\\
 & \\
\Lambda_{\rm ind} & = \frac{V(\varphi)}{\epsilon \varphi^2} \,\,.
\end{array}
\end{equation}

We will not assume a specific form for $V(\varphi)$, except that it is of the
induced gravity type, for example, $V(\varphi) = \frac{\lambda}{8} (\varphi^2
- \upsilon^2)^2$. Note that $\epsilon ,\,\, \lambda ,\,\,$ and $\upsilon$ are
all small constants.
For quantum cosmology, gravitation is always dominant, and we may neglect 
terms representing other fields in the action, Eq.\ref{action}.
Under the minisuperspace approximation, the metric of spacetime is given by
\begin{equation}
\label{metric}
ds^2 =  -N(t)^2 dt^2 + a(t)^2 d\Omega_3^2  \,\,,
\end{equation}
where $d\Omega_3^2$ is the line element of the three dimensional unit sphere,
$N(t)$ is the lapse function. 
The scalar gravitational field $\varphi$ depends on $t$ only. 
The total action can thus be written as 
\begin{equation}
\label{act}
S = \int dt \, 2\pi^2\left[ -3\epsilon\frac{a}{N} \varphi^2 \dot{a}^2
	- 6\epsilon\frac{a}{N} a\varphi\dot{a}\dot{\varphi}
	+ \frac{1}{2}\frac{a}{N} a^2 {\dot{\varphi}}^2
	+ 3\epsilon\frac{N}{a} a^2 \varphi^2
	-\frac{N}{a} a^4 V(\varphi) \right]  \,\,,
\end{equation}
where the dot stands for derivatives with respect to $t$.
The momenta conjugate to $a$ and $\varphi$ are defined in usual way 
and are respectively  given by
\begin{equation}
\label{momenta}
\begin{array}{ll}
\Pi_a & \equiv \frac{\delta S}{\delta \dot{a}} 
	= 2\pi^2\left( -6\epsilon\frac{a}{N} \varphi^2 \dot{a} 
	- 6\epsilon\frac{a}{N} a \varphi \dot{\varphi}\right) \,\,,\\
 &  \\
\Pi_{\varphi} & \equiv \frac{\delta S}{\delta \dot{\varphi}}
	=  2\pi^2\left( -6\epsilon\frac{a}{N} a \varphi \dot{a} 
	+ \frac{a}{N} a^2 \dot{\varphi} \right) \,\,.
\end{array}
\end{equation}
Then we have the following relations,
\begin{equation}
\label{relation}
\begin{array}{ll}
	a \dot{\varphi} = & \frac{N}{2\pi^2 a} 
	\frac{\varphi \Pi_{\varphi} - a \Pi_a}{(6\epsilon+1)a\varphi}\,\,, \\
 &  \\
	\dot{a} \varphi = & -\frac{N}{2\pi^2 a}
		\frac{6\epsilon\varphi\Pi_{\varphi} + a \Pi_a}{6\epsilon (6\epsilon+1)a\varphi} \,\,.
\end{array}
\end{equation}
Taking the variation of the action, Eq.\ref{act}, with respect to the lapse 
function $N$ and combining with Eq.\ref{relation}, we obtain the Hamiltonian 
constraint for the induced gravity theory,
\begin{equation}
\label{H}
H = - \left( 12\epsilon(6\epsilon + 1) a\varphi^2 \right)^{-1}
	\left({\Pi_a}^2 -6\epsilon\frac{\varphi^2}{a^2}{\Pi_{\varphi}}^2 
			+ 12\epsilon\frac{\varphi}{a} \Pi_a \Pi_{\varphi}\right)
	-\left(2\pi^2\right)^2 a [3\epsilon\varphi^2 - a^2V(\varphi)]
  = 0 \,\,.
\end{equation}
By introducing the canonical quantization into the above Hamiltonian constraint,
$\Pi_a \rightarrow \frac{1}{i} \frac{\partial}{\partial a}, \, 
\Pi_{\varphi}\rightarrow \frac{1}{i} \frac{\partial}{\partial{\varphi}}$,
we obtain the WDWE for the induced gravity theory,
\begin{equation}
\label{wdwe}
\left\{ \frac{\partial^2}{\partial a^2}
	- 6\epsilon \frac{{\varphi}^2}{a^2} \frac{\partial^2}{\partial {\varphi}^2}
	+ 12\epsilon \frac{\varphi}{a} \frac{\partial}{\partial a} \frac{\partial}{\partial \varphi}
	+ 48\pi^4\epsilon(6\epsilon+1) a^2 {\varphi}^2 [a^2V(\varphi)-3\epsilon\varphi^2] \right\} \Psi (a, \varphi) = 0 \,\,.
\end{equation}
In constructing the WDWE (Eq.\ref{wdwe}), we have ignored the operator order 
problem which is not important in the following discussion.

\section{WKB Wave Functions of the Universe}

In order to make predictions, we should solve the WDWE Eq.\ref{wdwe},
which proves to be a very difficult task. 
A regulous solution seems difficult, so we look for some simple solutions
that do not depend sensitively on $\varphi$.
We neglect the second and third terms in Eq.\ref{wdwe} to get
cosmic wave functions.
The problem is then simplified to a standard one-dimensional WKB problem
for the scale factor $a$ with a potential
\begin{equation}
\label{potential}
U(a) = 48\pi^4\epsilon(6\epsilon+1) a^2 {\varphi}^2 [3\epsilon\varphi^2 - a^2V(\varphi)] \,,
\end{equation}
For the classicall allowed (oscillatory) region 
$a \geq a_{_H} \equiv \left[ 3\epsilon\varphi^2/V(\varphi) \right]^{1/2}$, 
where the scale factor is large, there are WKB solutions of the form
\begin{scriptsize}  
\begin{equation}
\label{Psi1}
\begin{array}{ll}
\Psi_\pm (a,\varphi) &
    =[p(a)]^{-1/2} \exp \left[ \pm i\int_{a_{_H}}^a p(a')da'\mp i\pi /4 \right]\\
             &
    =\left[ 4\pi^2\sqrt{3\epsilon(6\epsilon + 1)} a\varphi \sqrt{a^2V(\varphi) 
	- 3\epsilon\varphi^2} \right]^{-1/2} \exp \left[ \pm i
	\frac{4\pi^2\sqrt{3\epsilon(6\epsilon + 1)}\varphi}{3V(\varphi)}
	\left[a^2V(\varphi)- 3\epsilon\varphi^2\right]^{3/2} \mp i\pi /4\right]
	, \,\,\, a\geq a_{_H},
\end{array}
\end{equation}
\end{scriptsize}  
where $p(a)=[-U(a)]^{1/2}$. 
For the classically forbiden (exponential) region $a < a_{_H}$,
where the scale factor is small, there are WKB solutions of the form
\begin{scriptsize}  
\begin{equation}
\label{Psi2}
\begin{array}{ll}
{\tilde \Psi}_\pm (a,\varphi)  &
	= |p(a)|^{-1/2}\exp\left[\pm\int_a^{a_{_H}}|p(a')|da'\right]\\
		&
	= \left[ 4\pi^2\sqrt{3\epsilon(6\epsilon + 1)} a\varphi 
       \sqrt{(3\epsilon\varphi^2 - a^2V(\varphi))}\right]^{-1/2} \exp \left[\pm 
	  \frac{4\pi^2\sqrt{3\epsilon(6\epsilon + 1)}\varphi}{3V(\varphi)}
	  \left[3\epsilon\varphi^2 - a^2V(\varphi)\right]^{3/2}	\right] 
	, \,\,\, a\le a_{_H}.
\end{array}
\end{equation}
\end{scriptsize}  
We can impose the boundary condition in either the classically allowed
region or the classically forbidden region, and then match the 
solutions in the two regions by the WKB standard matching procedure
to specify the WKB wave function.

There are several comprehensive and well studied boundary condition
proposals in the literature
(for a recent review, see Ref.~\cite{vil98a}).
Now we seek for the specified WKB wave function with different boundary
conditions following Vilenkin\cite{vil98a}.
Hartle \& Hawking\cite{har83} proposed that the specified wave function
should be given by the Euclidean path integral,
$\Psi_{HH} = \int e^{-S_E}$,
which is taken over compact Euclidean geometries and matter fields with 
a specified field configuration at the boundary.
Note that $S_E$ is the Euclidean action.
The Hartle-Hawking wave function is specified by requiring that it is
given by $\exp (-S_E)$ in the Euclidean regime. This gives\cite{har83}
\begin{equation}
\label{HH}
\begin{array}{ll}
\Psi_{HH}(a<a_{_H},\varphi)=& {\tilde \Psi}_-(a,\varphi),\\
\Psi_{HH}(a>a_{_H},\varphi)=& \Psi_+(a,\varphi)-\Psi_-(a,\varphi).
\end{array}
\end{equation}
However, Linde\cite{lin84} argued that the wave function should be given by,
$ \Psi_L = \int e^{+S_E}$,
which requires a reverse sign of the exponential in the Euclidean regime.
Together with the continuation to the classically allowed range of $a$,
one can get the Linde wave function as
\begin{equation}
\label{L}
\begin{array}{ll}
\Psi_L(a<a_{_H},\varphi)=& {\tilde \Psi}_+(a,\varphi),\\
\Psi_L(a>a_{_H},\varphi)=& {1\over{2}}[\Psi_+(a,\varphi)+\Psi_-(a,\varphi)].
\end{array}
\end{equation}

In addition, Vilenkin suggested that the wave function of the universe 
should be specified 
either by the tunneling boundary condition\cite{vil86} or by a Lorentzian 
path integral\cite{vil84},
$ \Psi_V = \int e^{iS}$.
Let us give a more precise statement of the Vilenkin's tunneling
boundary condition as follows\cite{hal91,pin96}:
$\Psi_V$ is the solution to the WDWE that is everywhere bounded and only
consists of outgoing modes at singular boundaries of superspace.
The superspace for our model is two-dimensional space with coordinates
($a, \varphi$), where, $0<a<\infty,\,\, -\infty < \varphi < \infty$.
The unique non-singular part of the boundary is $a=0$ with $\varphi$ 
being finite.  The rest are singular and consist of configuration with 
one or both $a$ and $\varphi$ being infinite. 
Note that as $a$ approaches zero, the coefficients of the second and third 
terms in Eq.\ref{wdwe} blow up. 
As the boundary condition requires, we are to get a regular solution, it seems 
reasonable to neglect the second and third terms in Eq.\ref{wdwe} to 
get the wave function of the universe.
It is easy to check that $\Psi_-(a,\varphi)$ and $\Psi_+(a,\varphi)$
describe an expanding and a contracting universe, respectively.
Vilenkin's boundary condition requires that only the expanding component
$\Psi_-(a,\varphi)$ should be present at large $a$.
The wave function within the quantum barrier can be found from the WKB 
connection formula.  Finally, we get the Vilenkin wave function,
\begin{equation}
\label{V}
\begin{array}{ll}
\Psi_V (a<a_{_H},\varphi)=& {\tilde \Psi}_+(a,\varphi)-{i\over{2}}{\tilde\Psi}_-(a,\varphi)\,,\\
\Psi_V (a>a_{_H},\varphi)=& \Psi_-(a,\varphi)\,.
\end{array}
\end{equation}
This completes the calculation of the wave function of the universe under
different types of boundary condition.

	\section{Result and Discussion}

Unfortunately, there has been no consensus so far on the interpretation of the
wave function of the universe among the quantum cosmology community, except
the statement that when $\Psi \sim \exp [iS/\hbar]$, classical behavior
should be recovered\cite{kol90}.
In order to understand the physical meaning of the resulting wave functions, 
we employ the so-called ``peak interpretation'', in which a prediction is 
said to be made when the wave function is sharply peaked in a certain region 
and almost zero elsewhere\cite{hal87,pin97}.
It is worthwhile to point out that, the causal or the the Bohm-de Broglie 
interpretation will lead to the same result\cite{pin97}.

For the classical region, $a\geq a_{_H}$, where spacetime has the
classical meaning and the classical solutions are valid, all three kinds 
of wave function are essentially oscillating, so the probability
distribution does not depend sensitively on $a$ or $\varphi$.
This independence implies that there is approximatively equal probability
for each point on the classical trajectories.
Therefore, the properties of the wave function of the universe in the
quantum era is crucial for the subsequent evolution of the universe.
Within the quantum barrier, both Linde's and Vilenkin's wave functions
are dominated by the decaying exponential ${\tilde\Psi}_+(a,\varphi)$,
which piles up at $V(\varphi) = 0$, i.e., $\varphi^2=\upsilon^2$, in the 
case of $V(\varphi)=\frac{\lambda}{8}(\varphi^2-\upsilon^2)^2$.
Since $\epsilon$ is small, the distribution will be concentrated in the 
narrow region around $\varphi^2=\upsilon^2$. 
Therefore the gravitational and cosmological ``constants'', $G_{\rm ind}$,
$\Lambda_{\rm ind}$, acquire constant values, and especially, the latter is
equal to zero.
It implies that no matter how the cosmological constant can vary in the 
classical models, it will posses zero value when the universe comes from 
quantum creation.
Nevertheless, Hartle-Hawking wave function within the barrier contain only the 
growing exponential ${\tilde\Psi}_-(a,\varphi)$, which piles up at 
$V(\varphi) = V_{max}$, where $V_{max}$ is the maximum value of $V(\varphi)$. 
Hence the universe would prefer a large cosmological constant.

In summary, we have investigated a quantum cosmological model  with the 
induced gravity theory.
After the WDWE was constructed in the minisuperspace approximation, we have
solved it using three kinds of boundary condition.
We have shown that the amplitude of the resulting tunneling wave function
sharply peaks around the classical trajectory only for a vanishing cosmological
constant.

\begin{acknowledgements}
We thank the anonymous refree for valuable comments, Prof. L. M. Krauss
for helpful suggestions and Prof. T. Kiang of Dunsink Observatory, Ireland, 
for polishing up the English. 
This work was supported by the National Natural Science Foundation of China,
under Grant No. 19903002.
\end{acknowledgements}



\begin{references}
\bibitem{bra61} C. Brans and R. Dicke, {\it Phys. Rev.} {\bf 124}, 925(1961).
\bibitem{zee79} A. Zee, {\it Phys.Rev.Lett.} {\bf 42}, 417(1979).
\bibitem{kra92gil97} L. M. Krauss and M. White, 1992, {\it Astrophys. J.}, {\bf 397}, 357(1992); G. T. Gillies, {\it Rep. Prog. Phys.} {\bf 60}, 151(1997).
\bibitem{tho96} S. E. Thorsett, {\it Phys.Rev.Lett.} {\bf 77} 1432(1996).
\bibitem{yos92} Y. Yoshii and K, Sato, {\it Astrophys. J.}, {\bf 387}, L7(1992).
\bibitem{blo96} L. F Bloomfield Torres and I. Waga, {\it MNRAS}, {\bf 279}, 712(1996).
\bibitem{via97} P. T. P. Viana and A. R. Liddle, {\it Phys. Rev.} {\bf D57}, 674(1998).
\bibitem{wei89} S. Weinberg, {\it Rev. Mod. Phys.} {\bf 61}, 1(1989).
\bibitem{car92} S. M. Carroll, W. H. Press and E. L. Turner, {\it Ann. Rev. Astron. Astrophys.} {\bf 30}, 499(1992).
\bibitem{hal91} J. J. Halliwell, ``An Introduction on Quantum Cosmology'', in: { \it Proceedings of Jerusalem Winter School on Quantum Cosmology and Baby Univers e}, ed. by S. Coleman, J. B. Hartle, T. Piran and S. Weinberg (Singapore: World Scientific, 1991).
\bibitem{zhu909298}  Z. H. Zhu, C. G. Huang and L. Liu, {\it Chin. Phys. Lett} {\bf 7}, 477(1990); Z. H. Zhu, {\it ibid.}  {\bf 9}, 273(1992); Z. H. Zhu, Y. Z. Zhang and X. P. Wu, {\it Mod.Phys.Lett.} {\bf A13}, 1333(1998).
\bibitem{mo88} H. J. Mo and L. Z. Fang, {\it Phys. Lett.} {\bf B201}, 321(1988).
\bibitem{har83} J. B. Hartle and S. W. Hawking, {\it Phys. Rev.} {\bf D28}, 2960
(1983).
\bibitem{lin84} A. D. Linde, {\it Lett. Nuovo Cimento} {\bf 39}, 401(1984).
\bibitem{vil86} A. Vilenkin, {\it Phys. Rev.} {\bf D33}, 3560(1986); {\bf D37} 888(1988).
\bibitem{vil98a} A. Vilenkin, {\it Phys. Rev.} {\bf D58}, 067301(1998).
\bibitem{vil84} A. Vilenkin, {\it Phys. Rev.} {\bf D30}, 509(1984).
\bibitem{pin96} N. Pinto-Neto, ``Quantum Cosmology'', in: {\it Cosmology and Gravitation II}, ed. by M. Novello (Editions Frontieres, Gyf-sur-Yvette, 1996).
\bibitem{kol90} E. W. Kolb and M. S. Turner, {\it The Early Universe}, p457 (Addison-Wesley Publishing Company, 1990).
\bibitem{hal87} J. J. Halliwell, {\it Phys. Rev.} {\bf D36}, 3626(1987).
\bibitem{pin97} J. Ac\'acio de Barros and N. Pinto-Neto, {\it Nucl. Phys.} {\bf B (Proc. Suppl.) 57}, 247(1997).
\end{references}
\end{document}